\documentclass[11pt]{article}
\usepackage[margin=1in]{geometry}
\usepackage{graphicx}
\usepackage{graphicx,epsfig}

\def\beq{\begin{equation}}
\def\eeq#1{\label{#1}\end{equation}}
\def\eeqn{\end{equation}}

\def\beqa{\begin{eqnarray}}
\def\eeqa#1{\label{#1}\end{eqnarray}}
\def\eeqan{\end{eqnarray}}

\let\bar=\overbar

\def\Dslash{\not{\hbox{\kern-4pt $D$}}}
\def\dslash{\not{\hbox{\kern-2pt $\del$}}}

\def\msb{{\bar{\ssstyle M \kern -1pt S}}}

\def\Title#1{\begin{center} {\Large {\bf #1} } \end{center}}
\def\Author#1{\begin{center} {\normalsize {\sc #1} } \end{center}}
\def\Institution#1{\begin{center} {\normalsize {\it #1} } \end{center}}
\def\Abstract#1{\noindent {\normalsize {\bf Abstract:} {\normalfont #1}}}
\def\Conference{\vspace{4mm}\begin{raggedright} {\normalsize {\it Talk presented at the 2019 Meeting of the Division of Particles and Fields of the American Physical Society (DPF2019), July 29--August 2, 2019, Northeastern University, Boston, C1907293.} } \end{raggedright}\vspace{4mm}}

\begin{document}

%
%

\Title{New results on the search for rare kaon events with the KOTO detector}

\Author{\underline{B. Beckford} on behalf of the KOTO collaboration}

\Institution{Department of Physics\\ University of Michigan, Ann Arbor, MI 48109 USA}

\Abstract{ 
The KOTO experiment was designed to observe and study the  K$^{0}_{L} \rightarrow$ $\pi^{0}\nu\bar{\nu}$ decay at J-PARC. The Standard Model (SM) prediction for the process is (3.0 $\pm$ 0.3) x 10$^{-11}$ with small uncertainties. This unique \emph{golden} decay is an ideal candidate to probe for new physics and can place strict constraints on beyond the standard model (BSM) theories. The previous experimental upper limit of the branching ratio was set by the KEK E391a collaboration as BR(K$^{0}_{L} \rightarrow \pi^{0}\nu\bar{\nu}$) $<$ 2.6 x 10$^{-8}$.
The signature of the decay is a pair of photons from the $\pi^{0}$ decay and no other detected particles. For the measurement of the energies and positions of the photons, KOTO uses a Cesium Iodide (CSI) electromagnetic calorimeter as the main detector, and hermetic veto counters to guarantee that there are no other detected particles. KOTO's initial data was collected in 2013 and achieved a similar sensitivity to the E391a result. We completed hardware upgrades and had the first major physics runs in 2015. This proceeding summarizes the presentation of recent results from the 2015 runs on the search for the decay K$^{0}_{L} \rightarrow \pi^{0}\nu\bar{\nu}$ with the data collected in 2015 in the KOTO experiment. The best experimental upper limit on the branching fraction from a direct search was set as  BR(K$^{0}_{L} \rightarrow \pi^{0}\nu\bar{\nu}$)$<$ 3 x 10$^{-9}$ at the 90\% confidence level. }

\Conference

%
%

\section{Introduction}
The K$^{0}_{L} \rightarrow \pi^{0}\nu\bar{\nu}$ decay is a highly suppressed process in the Standard Model due to the Flavor Changing Neutral Current (FCNC) s$\rightarrow$ d transition. This decay is forbidden at the tree-level and must proceed only by a loop process~\cite{Littenberg}. The KOTO experiment at the Japan Proton Accelerator Research Complex is dedicated to studying this rare decay. The model-independent  Grossman-Nir bound on the branching ratio limit of K$^{0}_{L} \rightarrow \pi^{0}\nu\bar{\nu}$ was set from isospin symmetry arguments based on the branching ratio of BR(K$^{+} \rightarrow \pi^{+}\nu\bar{\nu}$)  reported by the BNL E949 experiment~\cite{Artamonov}. This Grossman-Nir bound has constrained the branching fraction of the neutral decay to BR(K$^{0}_{L} \rightarrow \pi^{0}\nu\bar{\nu})  < 1.46$ x $10^{-9}$~\cite{Grossman}. Currently the SM decay rate prediction is (3.0 $\pm$ 0.309 $\pm$ 0.06) x 10$^{-11}$~\cite{Buras_2}. 

KOTO had its first physics run in 2013. Data was gathered from 100 hours of Protons on Target (POT). There was one event observed in the signal region with 0.34 $\pm$ 0.16 background events estimated from a blind analysis technique. Based on the observation of one event in the signal region, the collaboration placed upper limit of BR(K$^{0}_{L} \rightarrow \pi^{0}\nu\bar{\nu})< $5.1 x 10$^{-8}$  at the 90\% confidence level (C.L.)~\cite{Maeda}, which was comparable to the sensitivity and upper limit set by the KEK E391a experiment~\cite{Brod,Buras,Ahn}

\section{KOTO Experiment}
\subsection{KOTO detector}
\label{sec:Detector}
  \begin{figure}[htb]
	\begin{center}
	\epsfig{file=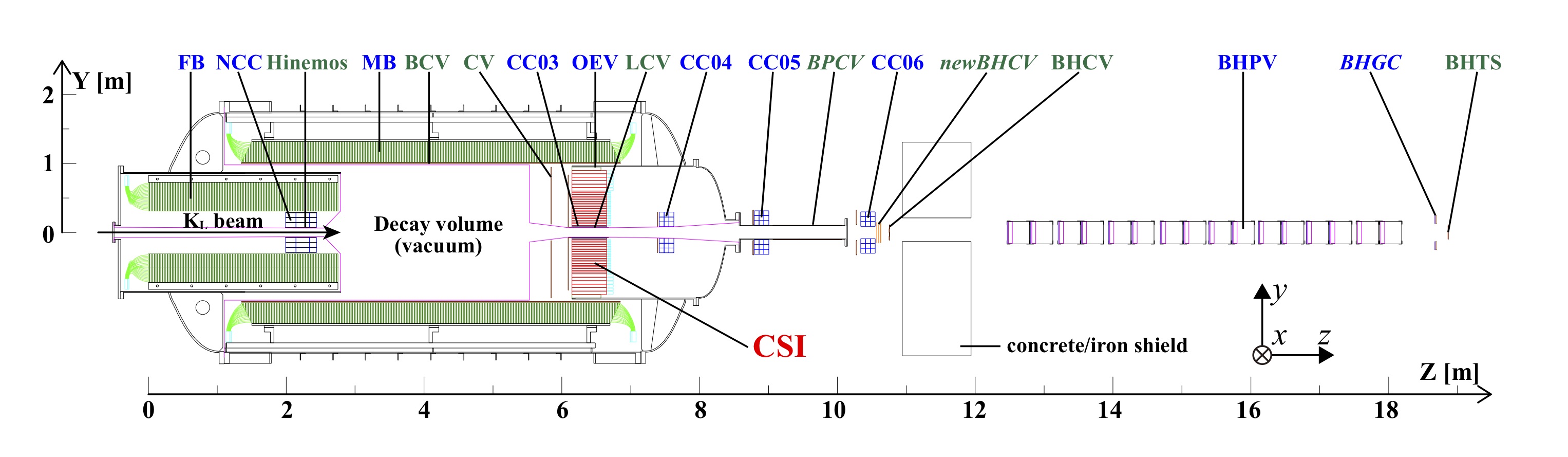,width=1.0\columnwidth}
			\caption{(Color online) Cross-sectional side view of the KOTO detector. The beam enters from the left side and goes along the z axis.}
  	\vspace{-.5cm}
	\label{fig:koto_schematic}
	\end{center}
\end{figure}

The KOTO experiment is performed at the J-PARC Hadron Experimental Facility (HEF) and aims to observe the rare Flavor Changing Neutral Current (FCNC) decay of K$^{0}_{L} \rightarrow \pi^{0}\nu\bar{\nu}$ decay by measuring $\pi^{0}\rightarrow \gamma\gamma$ and nothing else. Kaons were generated from a beam of 30-GeV protons that are slowly extracted  from the Main Ring (MR) with stable beam intensity. The proton beam struck a 66-mm-long gold target at HEF and produced a neutral beam that was directed at 16 degrees from the primary proton beam to the KOTO detector passing through a photon absorber, sweep magnet, and a pair of collimators. The extraction at this angle reduced the neutron energies that came as part of the neutral beam, and thus reducing background source for our measurement. The neutral beam was collimated to achieve a so-called pencil beam with a solid angle of 7.8 $\mu$sr and a size of 5 x 5 cm$^2$  downstream of the target. This size constraint was used as an additional restriction in determining the decay position in the transverse direction to the beam. \emph{Halo neutrons} were defined as neutrons that are part of the neutral beam but were located outside the nominal beam solid angle.  

A schematic view of the KOTO detector is shown in Fig.~\ref{fig:koto_schematic}. The origin of the z-axis, which lays along the beam-direction, was the upstream edge of Front Barrel detector (FB), located 21.5m away from the primary target. The x (horizontal) and y (vertical) axes were defined as standard in a right-handed coordinate system. The KOTO detector consisted of the CsI calorimeter (CSI) and hermetic veto counters surrounding the decay volume in vacuum. The CSI was composed of 2716 undoped CsI crystals whose length was 50 cm and cross-section was 2.5 x 2.5 cm$^2$ (5 x 5 cm$^2$) within (outside) the central 1.2 x 1.2 m$^2$ region. A 15 x 15 cm$^2$ center region of the stacked CSI permitted the beam particles to pass through the detector. The KOTO detector used lead-scintillator sandwich, lead-aerogel, or undoped CsI for the  photon veto counters, and plastic-scintillators or wire chambers as the charged particles vetoes.

\subsection{Experimental method}
The signature of the signal decay is characterized with two photons from the $\pi^{0}$  decay and nothing else. We used a hermetic veto system to ensure there are no other particle signals and thus meeting the "nothing else" requirement. The two photons were detected with theelectromagnetic calorimeter (CSI) in Fig.~\ref{fig:koto_schematic} and we ensured that there were no other particles in the veto counters. The vertex position was reconstructed assuming the nominal $\pi^{0}$ mass and a decay vertex (Zvtx) located on the beam axis. We then calculated the transverse momentum (Pt) and defined our signal area in the plane formed from the $\pi^{0}$ transverse momentum (Pt) and the decay vertex (Zvtx).

\section{Results}
 \begin{figure}[ht]
	\begin{center}
		\epsfig{file=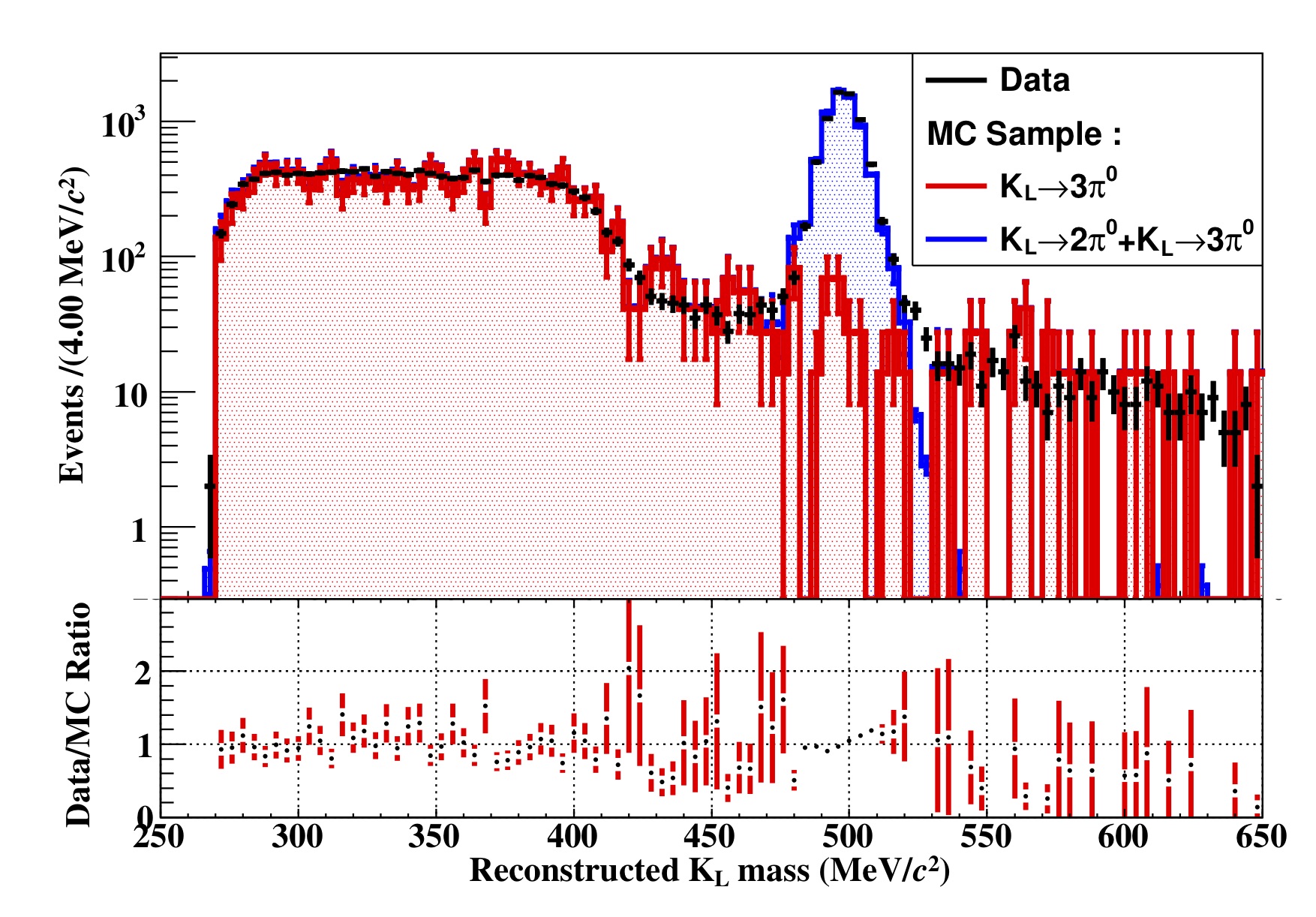,width=0.70\columnwidth}
			\caption{(Color online) Reconstructed  four photon invariant mass for K$^{0}_{L}\rightarrow \pi^0 \pi^0$ events with all the cuts. The ratio of data to MC (sum of  K$^{0}_{L}\rightarrow \pi^0 \pi^0$ and K$^{0}_{L}\rightarrow \pi^0 \pi^0\pi^0$) for each histogram bin is given in the bottom panel in the figure.}
			  \vspace{-.2cm}
	\label{fig:KL_mass}
	\end{center}
\end{figure}

The data taken during 2015 (Runs 62$-$65), at beam powers of 24-42 kW was proportional to  2.2 x $10^{19}$ protons on target. We also recorded data for normalization and calibration studies via the decays of K$^{0}_{L} \rightarrow \pi^{0}\pi^{0}\pi^{0}$, K$^{0}_{L} \rightarrow \pi^{0}\pi^{0}$, and K$^{0}_{L} \rightarrow \gamma\gamma$. Figure~\ref{fig:KL_mass} presents the reconstructed K$^{0}_{L}$  mass distribution with all selections applied except the requirement that events be between $\pm$15 MeV/c$^2$ of the mass peak. 

We classified backgrounds into two distinct source groups: K$^{0}_{L}$decays and background originating from neutron induced backgrounds caused by halo-neutrons hitting a detector component~\cite{Naito}. The backgrounds from K$^{0}_{L}$ decays was estimated using MC simulations. Background estimations were performed while masking the data in the region 2900 $<$ Zvtx $<$ 5100 mm and 120$<$ Pt$<$ 260 MeV/c to avoid introducing bias. Based on MC acceptance studies and the number of reconstructed K$^{0}_{L} \rightarrow \pi^{0}\pi^{0}$ events in the data set, after applying all cut selections, we obtained a single event sensitivity for the  K$^{0}_{L} \rightarrow \pi^{0}\nu\bar{\nu}$ decay as (1.30$\pm$0.01$_{stat.}\pm$0.14$_{syst.}$) x 10$^{-9}$.

 \begin{figure}[ht]
	\begin{center}
		\epsfig{file=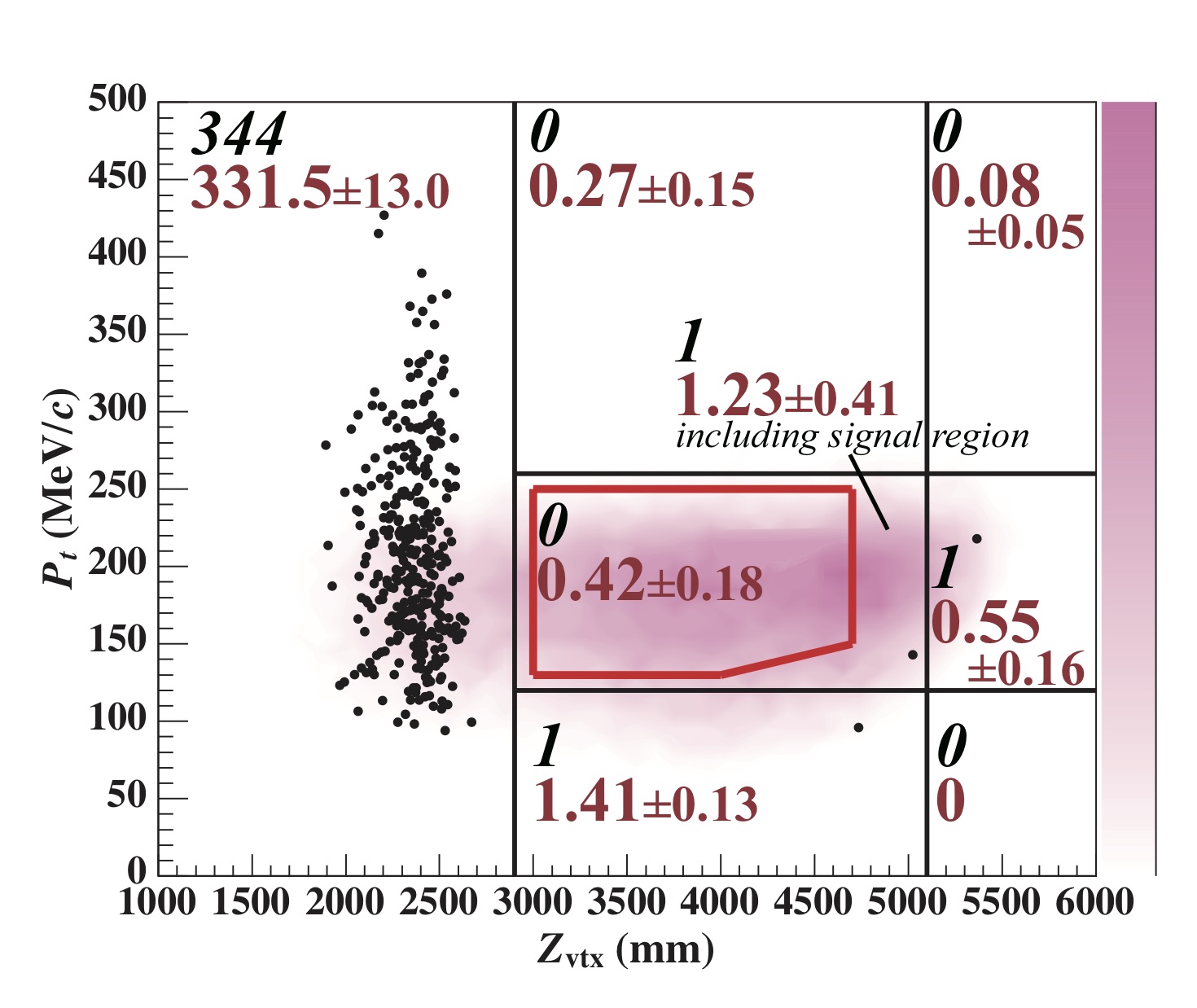,width=0.700\columnwidth}
			\caption{(Color online) Reconstructed $\pi^0$ transverse momentum (Pt) versus reconstructed decay vertex (Zvtx) from the 2015 data set with all cuts applied. The red lines indicate the signal area for K$^{0}_{L} \rightarrow \pi^{0}\nu\bar{\nu}$. The observed events for data and expectations from Monte-Carlo in different regions are shown as the black and red numbers, respectively.}	
 	\vspace{-.2cm}
	\label{fig:2015_results}
		\end{center}
\end{figure}

Figure~\ref{fig:2015_results} presents the reconstructed $\pi^0$ transverse momentum (P$_t$) versus decay vertex (Zvtx) for the 2015 data set with all cuts applied. The signal candidates were required to be in the region determined by 130$<$ Pt$<$ 260 MeV/c,  3000 $<$ Zvtx $<$ 4000 mm, and varied linearly from 130 - 150 MeV/c in the range from 4000 $<$ Zvtx $<$ 4700 mm, indicated by the red lines in Fig~\ref{fig:2015_results}. The Monte Carlo determined signal distribution is shown as the pink contour. The black (red) numbers give the numbers of observed (expected background) events for the each region. The events observed in both data and the MC estimations agreed well.
The total number of expected background events was (0.42 $\pm$ 0.18), where the major source originated from neutron events hitting the CSI (hadron cluster background). It was estimated from a control data sample taken by inserting an aluminum target into the neutral beam to scatter out the core neutrons as (0.24 $\pm$ 0.17). The summarized background contributions and sources are provided in Table~\ref{table:background_table_2015}. There was no candidate event observed in the signal region. Based on no observed candidate event, we employed Poisson statistics and the methodology outlined by Cousins et. al.~\cite{Cousins} and set an upper limit on the decay of BR(K$^{0}_{L} \rightarrow \pi^{0}\nu\bar{\nu})$ $<$ 3.0 x 10$^{-9}$~\cite{KOTO2015}.  

\begin{table}[h!]
\caption{Summary of estimated background estimation in the signal region}
\begin{center}
\label{table:background_table_2015}
\begin{tabular}{|l|c|}
  \hline
  Source of background & estimated events \\
  \hline
	K$^{0}_{L}\rightarrow \pi^{0}\pi^{0}$			& 0.02 $\pm$ 0.02 \\
	K$^{0}_{L}\rightarrow \pi^{0}\pi^{+}\pi^{-}$	& 0.05 $\pm$ 0.02 \\
	other K$^{0}_{L}$ decays						& 0.03 $\pm$ 0.01 \\
	Neutrons events on NCC (upstream)			& 0.04 $\pm$ 0.03 \\
	Halo neutron events hitting CsI				& 0.24 $\pm$ 0.17 \\
	 CV $\eta$									& 0.04 $\pm$ 0.02 \\

  	\hline
   	\hline
   	Total	& 0.42 $\pm$ 0.18	\\
	\hline
	\hline
\end{tabular}
\end{center}
\end{table}

\section{Upgrades and outlook }
The main background contribution in the signal region came from halo induced neutron reactions that could be mistaken for signal events. Primarily, this occurred when a neutron directly hit the CSI and induced showers that appeared as two photon events. A detector upgrade to reduce this background  involved the installation of Multi Pixel Photon Counters (MPPC) to the front of the CSI. We installed 4080 MPPCs in 2018 that will enable distinguishing between $\gamma$ and neutron events using timing differences from a both-end readout approach. A new trigger hardware component was installed to permit do finding and also to calculate the number of clusters at the online trigger level, aimed at improving the data acquisition (DAQ) efficiency. In order to reduce background events from K$^{0}_{L}  \rightarrow \pi^{0} \pi^{0}$ we installed a new barrel detector (IB) inside the Main Barrel that added an additional 5X$_0$. We also included new analysis procedures that improved the performance of the cluster shape discrimination for $\gamma$ and  neutron events by introducing deep learning using the energy and timing information of the clusters, and improved the pulse shape discrimination method with a Fourier transformation of the waveforms.
\subsection{Outlook}
We have successfully performed a search for K$^{0}_{L} \rightarrow \pi^{0}\nu\bar{\nu}$ with the KOTO detector with data taken in 2015 (Runs 62--65).  We established a new experimental an upper limit of BR(K$^{0}_{L} \rightarrow \pi^{0}\nu\bar{\nu})$ $<$ 3.0 x 10$^{-9}$ at a 90\% confidence level (C.L.), improving on the previous limit by an order of magnitude. The KOTO experiment will continue to search for the K$^{0}_{L} \rightarrow \pi^{0}\nu\bar{\nu}$ with data gathered in 2016-2018, and 2019.

\section*{Acknowledgements}
This work was supported by the U.S Department of Energy, Office of Science, under research Awards No. DE-SC0007859, DE-SC0006497, and DE-SC0009798.

\end{document}